\documentclass[sigconf]{acmart}
\usepackage{csquotes}
\usepackage{hyperref}
\AtBeginDocument{%
  \providecommand\BibTeX{{%
    \normalfont B\kern-0.5em{\scshape i\kern-0.25em b}\kern-0.8em\TeX}}}

\copyrightyear{2022}
\acmYear{2022}
\setcopyright{rightsretained}
\acmConference[UIST '22]{The 35th Annual ACM Symposium on User Interface Software and Technology}{October 29-November 2, 2022}{Bend, OR, USA}
\acmBooktitle{The 35th Annual ACM Symposium on User Interface Software and Technology (UIST '22), October 29-November 2, 2022, Bend, OR, USA}
\acmDOI{10.1145/3526113.3545693}
\acmISBN{978-1-4503-9320-1/22/10}

\acmJournal{JACM}
\acmVolume{37}
\acmNumber{4}
\acmArticle{6}
\acmMonth{10}

\usepackage{float}
\usepackage{array}
\usepackage{ctable} 
\usepackage{longtable}
\usepackage{multirow}

\begin{document}

\title{Fuse: In-Situ Sensemaking Support in the Browser}

\author{Andrew Kuznetsov}
\affiliation{
  \institution{Carnegie Mellon University}
  \city{Pittsburgh}
  \state{PA}
  \country{United States}
}
\email{adkuznet@cs.cmu.edu}

\author{Joseph Chee Chang}
\affiliation{
  \institution{Allen Institute for AI}
  \city{Seattle}
  \state{WA}
  \country{United States}
}
\email{josephc@allenai.org}

\author{Nathan Hahn}
\affiliation{
  \institution{US Army Research Labs}
  \city{Adelphi}
  \state{MD}
  \country{United States}
}
\email{nphahn@gmail.com}

\author{Napol Rachatasumrit}
\affiliation{
  \institution{Carnegie Mellon University}
  \city{Pittsburgh}
  \state{PA}
  \country{United States}
}
\email{nrachata@cs.cmu.edu}

\author{Bradley Breneisen}
\affiliation{
  \institution{Carnegie Mellon University}
  \city{Pittsburgh}
  \state{PA}
  \country{United States}
}
\email{bbreneis@cs.cmu.edu}

\author{Julina Coupland}
\affiliation{
  \institution{Carnegie Mellon University}
  \city{Pittsburgh}
  \state{PA}
  \country{United States}
}
\email{jgolze@cs.cmu.edu}

\author{Aniket Kittur}
\affiliation{
  \institution{Carnegie Mellon University}
  \city{Pittsburgh}
  \state{PA}
  \country{United States}
}
\email{nkittur@cs.cmu.edu}

\renewcommand{\shortauthors}{Kuznetsov, et al.}

\begin{abstract}


People spend a significant amount of time trying to make sense of the internet, collecting content from a variety of sources and organizing it to make decisions and achieve their goals. While humans are able to fluidly iterate on collecting and organizing information in their minds, existing tools and approaches introduce significant friction into the process. We introduce Fuse, a browser extension that externalizes users’ working memory by combining low-cost collection with lightweight organization of content in a compact card-based sidebar that is always available. Fuse helps users simultaneously extract key web content and structure it in a lightweight and visual way. We discuss how these affordances help users externalize more of their mental model into the system (e.g., saving, annotating, and structuring items) and support fast reviewing and resumption of task contexts. Our 22-month public deployment and follow-up interviews provide longitudinal insights into the structuring behaviors of real-world users conducting information foraging tasks.


\end{abstract}




\maketitle

\section{Introduction}

People spend a significant amount of time trying to make sense of the internet, collecting content from a variety of sources and organizing it to make decisions and achieve their goals. For example, consumers purchasing a home appliance spend an average of 15.8 hours researching across 12 different sources. \footnote{(Google and Ipsos MediaCT, January 2014) and (Google and Inmar, 2013) respectively, as referenced by The Google Marketing Blogpost "Thinking with Google: Five Holiday Shopping Trends Marketers Should Watch".} Similar sensemaking takes place for a variety of information seeking and decision tasks, ranging from shopping to health to travel to troubleshooting \cite{kittur2013costs}.

\begin{figure*}
  \includegraphics[width=\textwidth]{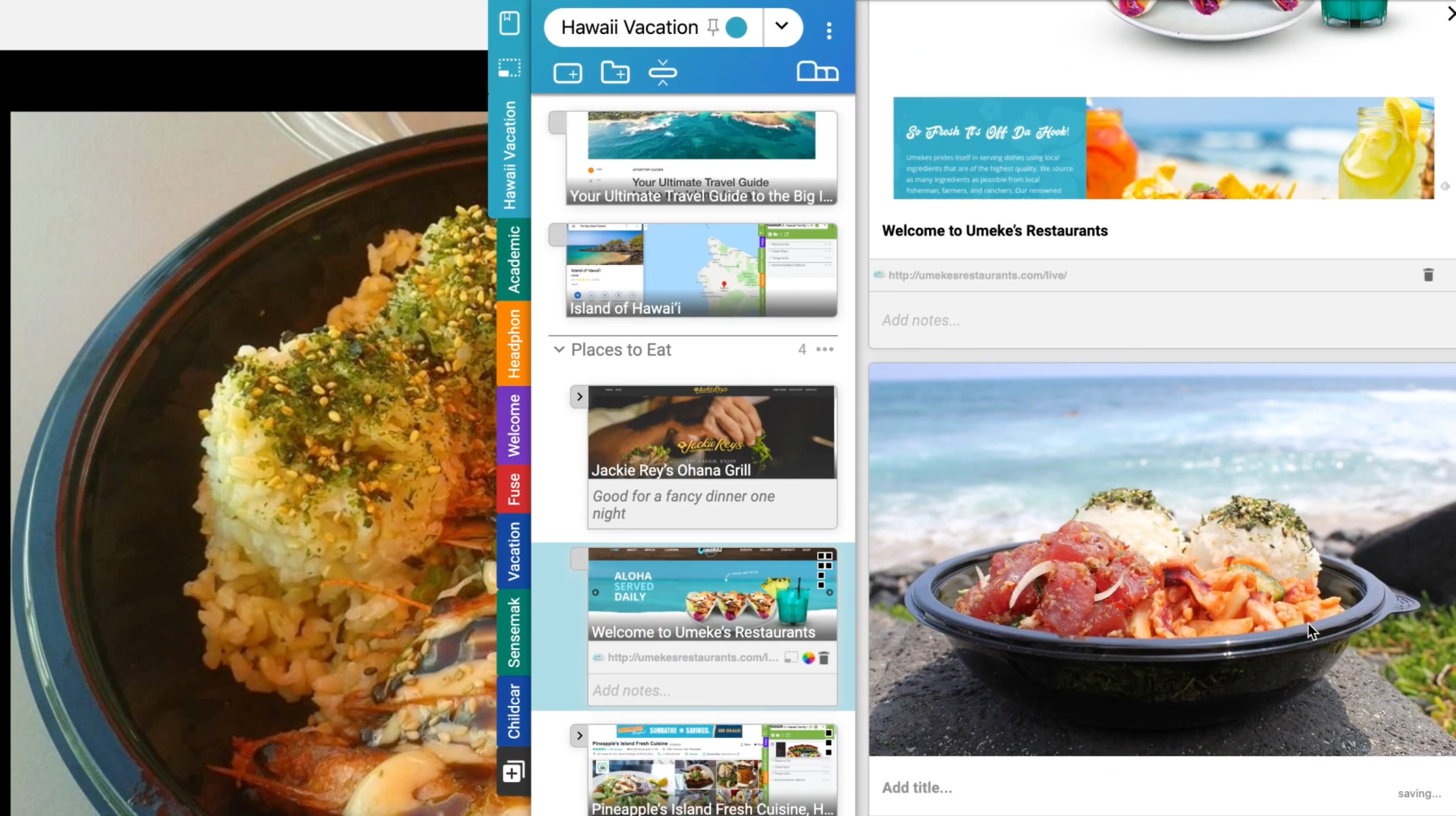}
  \vspace{-4mm}
  \caption{The Fuse Prototype, expanded to demonstrate how the reader view enables users to drill-down on specific items while maintaining a compact representation during foraging.}
  \Description{caption}
  \label{fig:teaser}
  \vspace{-2mm}
\end{figure*}

While humans are able to fluidly iterate on collecting and organizing information in their minds \cite{russell1993cost}, the amount of information, options, and evidence involved in online sensemaking quickly exceeds the limits of working memory \cite{Miller1960InformationIO}. However, externalizing these processes using existing tools and approaches introduces significant friction.  For example, a common way to collect information today is to use browser tabs to externalize and keep track of information. However, tabs' flat structure and lack of support for more complex organization leads to a multitude of issues, from tab overload \cite{chang2021tab} and tab hoarding \cite{vitale2018hoarding}\cite{sweeten2018digital}\cite{dubroy2010study} to losing users' mental context which isn't externalized anywhere \cite{hwang2017link}. These issues are not easily fixed by introducing tab groups or hierarchies which can correspondingly lead to tab group overload \cite{chang2021tabsdo}.

Exacerbating the problem, many online sensemaking tasks require extracting information scattered across pages and reassembling them into the concepts or items a user is considering \cite{Hoch1986ConsumerLA} \cite{Racherla2012PerceivedO} \cite{chang2020mesh}
\cite{Hahn2016TheKA} \cite{Chang2016AlloyCW} \cite{Chen2015TripPlannerPT}. For example, in the home appliance example above, a user may need to collect unique features or important limitations which may be discussed in different sources or reviews. Collecting and organizing this information by copy-pasting and switching between web pages and docs or spreadsheets can be onerous and time consuming \cite{chang2021tab}. As a result, tools that assist the user in clipping pieces or entire web pages have gained in popularity as an effort at combating these costs (e.g., as found in Evernote, hypothes.is, Pinterest, etc.). 


However, while the above approaches have focused on reducing the friction of collecting web content, users typically still need to switch to another application in order to access and organize that content. This separation between collection and organization is in contrast to the fluid and iterative sensemaking process users engage in when information can be contained entirely in their head. In this work we investigate the idea of providing users with an in-situ workspace that more fluidly bridges the gap between browsing, collecting, and organizing web content. Our goal is to take a step towards an interaction paradigm in which the browser acts as a seamless, zero-friction extension of a user's mental workspace.



There are many benefits that a browser acting as a seamless mental workspace could provide: quick access, resumption of tasks, recall of information origin and purpose, and scaffolding users in building larger and more complex information collections. In exploring this vision we grapple with several core design challenges. First, how do we reduce the friction of collecting information while keeping users in the flow of their sensemaking process? This goal is challenging because users may wish to collect a variety of types of information ranging from text to image to mixed-information clips and have access to both the final visual rendering of that information as well as the underlying text, links, and structures, all while maintaining the provenance of where those clips came from. Furthermore, as the user is reading and deciding to collect the information they are making judgments in their mind that need to be externalized as well, such as through notes and annotations such as the priority or expected utility of the item.

Second, how can we help users not only collect information in situ but also view and organize the information collected in situ as well? Requiring the user to switch between their information collection and another tab or app can reduce the likelihood of their actually using and structuring their information. Although previous systems have explored the value of capturing users' thoughts in an in-situ sidebar (e.g., \cite{van2009note}), less work has examined the ability to support informal and flexible organization in an always-available sidebar. Exploring the affordances of structure and organization needed by users was a key goal for this work that drove our approach of a long term deployment with completely voluntary usage.

Third, how do we help users visually represent and make sense of all the clips, pages, and notes collected and organized in the system? Given the limited real estate afforded by a sidebar, exploring compact representations that could compress and display a variety of diverse information collected was essential to the success of the approach. In doing so we explore flexible compression approaches such as 'container' cards that support compressing multiple pieces of evidence into a single card; visual previews for such containers; and an expanded 'reader' view which flattens clips, containers, and nested structures so that users can quickly scroll through them to get an overview of what they've collected, all without leaving the webpage they're on.



In this paper we present Fuse, an exploration of the idea of an externalized, unified workspace in the browser, and a prototype instantiating that idea that reduces the friction of collecting and organizing web content in the browser. We discuss the design and development of Fuse as conducted over the course of a year of iterative development by the research team and a small group of users using it for their own tasks, as well as a 22 month public deployment with over 100 volunteer users. Our findings validate the importance of supporting in-situ organization in the browser and identify several challenges for future systems aiming to support online sensemaking and task management.

Our contributions include:
\begin{itemize}

\item Fuse, a prototype browser extension that provides a unified workspace for conducting online research combining affordances for low-cost collection with lightweight organization of content in a compact card-based sidebar that is always available.
\item A analysis of how these affordances help users externalize their mental structures into the workspace created by the system (e.g., saving, annotating, and structuring items) and support fast reviewing and resumption of task contexts. 
\item A 22-month public deployment and follow-up interviews that provide longitudinal insights into the structuring behaviors of real-world users conducting information foraging tasks.

\end{itemize}


\section{Related Work}

\subsection{Clipping Content on the Web}
The larger topic of capturing and clipping web content is well studied in both industry and academia. By necessity, organization tools such as Notion \footnote{https://www.notion.so/}, Evernote \footnote{https://evernote.com/}, and Pocket \footnote{https://getpocket.com/} implement clipping tools at the DOM level to import web content to be organized within their respective applications. Clipping systems in academic research have previously focused on extracting specific media types such as images \cite{pham2012clui}, or text \cite{tashman2011liquidtext} \cite{kittur2014standing} \cite{liu_crystalline_2022} \cite{liu_wigglite_2022} \cite{Chang2016SupportingMS}, although clipping tools that leverage HTML structure also appear in several tools \cite{dontcheva2006collecting} \cite{schraefel2002hunter}. Another approach to extracting web content utilizes structured extraction, inducing a schema from an initial collection and using this to extract similar content \cite{hogue2005thresher} \cite{dontcheva2006collecting}, or repeatedly extracting updated content \cite{sugiura1998internet}. Fuse improves on existing clipping tools by forgoing extraction templates and instead making it easy to collect text, images, or screenshots of information by providing users with a variety of approaches to import content at multiple granularities, including drag and drop, screenshot clipping (which saves both image- and html-based representations), and bookmarking entire pages. Additionally, Fuse provides affordances for users to explicitly encode the context of information after importing it, for example assigning tags, colors, adding notes, or changing the order of items. Simultaneously, Fuse also automatically collects high-level provenance information to help users recover the original context of the content, such as origin url and a viewport screenshot.

\subsection{In-Situ Sensemaking Support Systems}
Researchers have explored a number of systems for supporting the sensemaking activities of collecting and organizing in different contexts such as literature review \cite{zhang2008citesense}, web search interfaces \cite{hearst2013sewing} \cite{Hahn2018BentoBC}, as well as desktop applications that mediate or review dedicated workspaces \cite{jeuris2014laevo} or review screen context \cite{hu2020screentrack}. Such in-situ systems are frequently implemented as standalone desktop programs that replace the user's typical browser experience, often consistent of a main reading panel and a working space for viewing or organizing collected content. A few notable examples include CiteSense \cite{zhang2008citesense}, which demonstrates this in the domain of literature review, allowing users to take notes on search results while querying a database in a standalone Java application. Hearst et al. \cite{hearst2013sewing} similarly explore the design of an interface for tagging and organizing search results, with the main work space consisting of a card-based clustering interface for highlighted documents. Reimer et al. \cite{reimer2011turning} introduced a desktop application that imports elements via copy and paste and splits organization into a two panels; a hierarchical file structure on the left-hand side enables users to expand an item to view on the right-hand side. However, the unstructured nature of web content mandates that respective sensemaking support systems must implemented interfaces that smoothly enable users to browse web content, extract it, and later review extracted web DOM content, whatever it may be.\looseness=-1


To better adapt such systems to the web content encountered while browsing the web, researchers have also explored in-situ ‘sidebar’ systems implemented as browser extensions. For example, systems like List-it \cite{van2009note} allow users to type free form notes and todos into a sidebar while browsing the web, while systems with built-in clippers such as Clipper \cite{kittur2013costs}, ForSense \cite{rachatasumrit2021forsense}, or Threddy \cite{threddy} allow users to collect and organize text snippets manually extracted from the web pages \cite{kittur2013costs}\cite{rachatasumrit2021forsense} or scholarly articles \cite{threddy}. These systems make it easy for users to preserve context across pages and quickly reaccess information while browsing, but have limited support for organization (typically at most one level of grouping) and require switching to another tab or application for viewing and interacting with collected content. As a result, many such systems end up visually representing clips in relatively unordered structures such as collages \cite{swearngin2021scraps}, with no structure or nesting possible. Systems that do arrange clippings with structure, such as Dontcheva et al's "Summary Tables" \cite{dontcheva2006collecting} require the creation of custom templates and views by users.

In Fuse we attempt to fuse the benefits of in-situ browser sidebars regarding situational awareness and rapid reaccess of information with the benefits of the organizations and visual structures found in dedicated workspaces. In order to support the user in doing so without context switching we explore approaches integrating user encoding of mental context, flexible organization, and visual compression.\looseness=-1

\section{System Design}

\subsection{Design Goals and Process}

We developed the Fuse protoype to probe the gap between collecting and organizing information online, which to date have been addressed by researchers as largely separate activities. As such our high level research goals were to provide a flexible set of affordances that could be used across a variety of online research tasks to reduce the friction of not only collecting web content but also organizing it in-situ and quickly reaccessing and using that organization. Conceptually, our design goal could be framed as trying to reduce the gap between the internal, evolving representations in the mind of the users as they collect online information, and the externalized, existing structure of their collection, eventually resulting in the browser functioning as a seamless extension of the user's mental workspace. This work represents a first step towards that goal.

With the above in mind, we developed Fuse to assist users in conducting online research by enabling them to collect and organize web document clips and links in-situ in a persistent browser sidebar. Typically, a user synthesizing online content would need to switch between the content they are exploring and a reference document with their collected items as they synthesize information across multiple online sources - a process often done with high context-switching costs using copy-and-paste and an external spreadsheet application \cite{chang2020mesh}. Existing systems use a combination of techniques for avoiding this issue, from `importing' entire tabs in Tabs.do \cite{Chang2021TabsdoTB} to automatically extracting pre-defined content (such as price and average reviews) from tabs in Mesh \cite{chang2020mesh}. Instead, our system enables users to reference and organize their collection of snippets as they simultaneously forage for more information. An inherent challenge to creating such a general purpose system includes supporting a variety of foraging and sensemaking activities that users undertake throughout a spectrum of domains and tasks. To enable such flexibility, Fuse combines paradigms such as web clipping, bundling, file management, and card interfaces to create a compact interface for information foraging. For example, ‘snipping’ a piece of content in the browser window creates a Fuse ‘card’ in the sidebar, which can then be organized alongside, in a hierarchical folder structure, or bundled as part of an existing card. As a complete system, Fuse is designed to afford users the flexibility to synthesize a wide variety of web content into whichever structure is most convenient for a given user's task and domain. 

To thoroughly explore the design space and produce a research prototype robust enough for a large-scale field deployment, three of the authors spent a year developing the extension while all team members used the extensions for a variety of online sensemaking tasks including learning, shopping, trip planning, literature review, UI design survey, market research, coursework management, and more. As we deployed Fuse to more users we conducted informal user studies throughout the development process to iterate on its design. For example, the early versions of Fuse only provided a flat list of cards within each project. We later introduced the ability to create folders to build hierarchical structures so that users who were planning trips with Fuse can separate their clips for restaurants, hotels, and places to visit but at the same time see all collected information within the same project view. In later iterations we further introduced the ability to put cards within cards because users were collecting multiple snippets for the same items which lead to long hierarchy structures that can be difficult to overview in the sidebar. Instead, the ability to create cards-in-cards allows them to curate a single card that contains all clippings for the same items (e.g., all the information about the same restaurant), making it easier for users to see all their choices in the same view for decision-making while enabling them to drill down and inspect each of their options in the reader view.

In the below sections we describe an example user experience, then discuss in more depth the various design challenges and decisions made during the iterative design and deployment of Fuse, broken roughly into three functions:


\begin{itemize}
    \item (D1) Collecting content while encoding provenance and context
    \item (D2) In-situ organizing
    \item (D3) Visually compressing items 
\end{itemize}

\begin{figure*}[h]
\centering
\includegraphics[width=\textwidth]{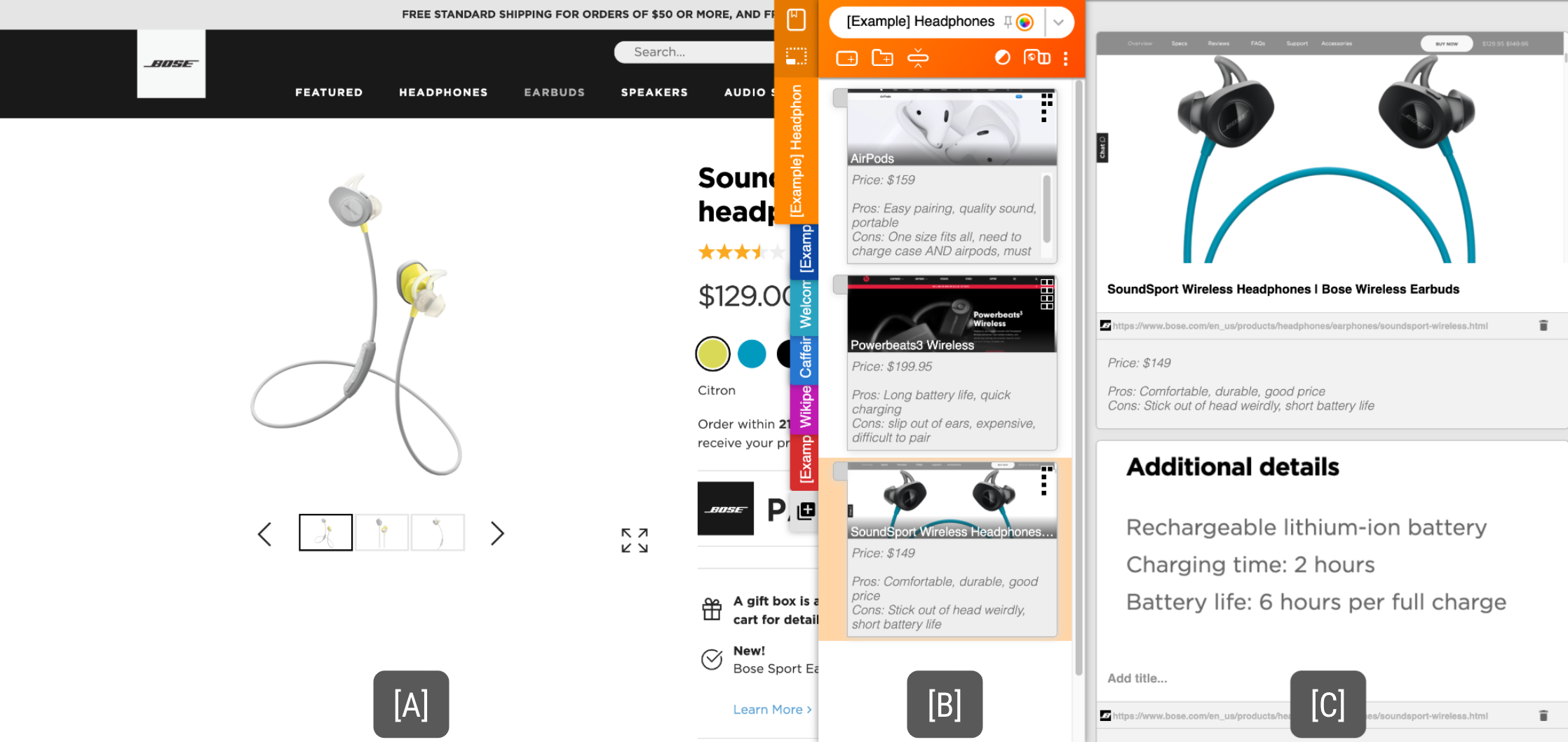}
\caption{The main interface of Fuse that is accessible to users alongside their web browser. (A) The web content, which can be freely interacted with as normal. (B) The main sidebar view for organizing collected items and selecting clipping interactions. In the screenshot, it has been expanded via clicking on one of the cards, revealing (C) the expanded reader view which enables users to dive deep into a specific card and view nested items in a compact manner. 
}
\end{figure*}

\subsection{Example User Experience}

Consider an example in which a user needs to conduct online research to find the best  pair of wireless earbuds for her needs. She starts by searching on Google for popular headphone models to consider, but quickly realizes that there are dozens of “best wireless headphones” listicles recommending hundreds of headphone models to track in her exploration. These listicles often contain overlapping recommendations, but might say different things about the same models.  Even collecting exclusively from a recommended list of headphones, she must cross-reference information about dozens of models from a variety of sources, such as Amazon reviews and other listicles. To understand which headphones might be best for her, she’d like to research the current headphones offerings and collect their respective cost, important features, and things-to-avoid, Normally, this process would typically take her a lot of effort, involving several hours of constantly switching between web pages, searches, and a separate external document such as a word document or a spreadsheet to use as external workspace keep track of her progress.  Instead, she uses Fuse to create a new project called “Wireless headphone shopping” and imports the different headphone models she found interesting through a variety of interactions: bookmarking entire product pages, extracting one-off recommendations from social media comments, or clipping a part of a listicle. The system then creates cards for each item and automatically recovers contextual information about the source, such as url, favicon, and screenshot of the viewport (except for cards created from direct clippings). Skimming over her list of cards, she reorders them by personal preference by dragging the cards relative to one another. 

As she dives deeper into researching specific options, she finds parts of pages that she’d want to clip and organize on product pages, such as impressive features and critical reviews. Using the Fuse sidebar, she quickly clips a review about one of her options and drags the newly created card into the option’s card, nesting it. Fuse creates a condensed representation of the nested cards while hiding them, letting her know the rough size of her bundle for each item. Although Fuse hides these nested cards for easier skimming, she can expand to the toolbar at any time to see all the clips collected under any card. As she accumulates more evidence and notes, she also creates folders to organize different types of wireless headphones (such as over-the-ear or earbuds), allowing her to externalize her task mental model and fluidly compress or expand the subtopic folders she is currently focusing her research on while maintaining a sense of situational awareness of the overall project. Finally, she feels confident and informed and makes a purchase with the information she has explored and saved in her Fuse project.


\subsubsection{[D1] Collecting content while encoding provenance and context} 

As illustrated in the example above, our first design goal was to allow those conducting online research to collect information while having direct access to their prior sensemaking collection. Prior work has demonstrated the utility of synthesizing information from multiple sources during sensemaking activities \cite{dontcheva2006collecting}. A system built to collect web content must enable users to quickly and effortlessly collect a diversity of web content (images, text, part of pages, videos, pdfs, etc.) as well as automatically capture the relevant surrounding context. The Fuse clipper collects content and context in a variety of representations. Firstly, a screenshot of the selected region is automatically saved as an image, preserving the visual appearance and styling and used as the default representation showing on cards to help users recall what they saved. Secondly, Fuse resolves the selected bounding region to the best-match Web DOM element and saves its HTML content. This allows users to access hyperlinks or select and copy text in the selected region. For PDFs opened in the browser, since PDFs do not contain DOM elements that contain blocks of text, we ran the Tesseract.js \footnote{https://tesseract.projectnaptha.com/} Optical character recognition library locally in the browser to convert the saved image into selectable text. Once saved, users can toggle between different automatically-collected representations in the sidebar based on their needs. When more context needs to be manually added, Fuse also implements text annotations that the user may add for any item. 

A critical design goal was for the collection interactions themselves to feel sufficiently lightweight for the user to allow users to seamlessly transition between browsing information and using Fuse to import content into their collection without breaking their flow. To achieve this, the Fuse clipper provides five interaction types that can be utilized by users to address different clipping needs (Figure 3):

\begin{figure*}[h]
\centering
\includegraphics[width=\textwidth]{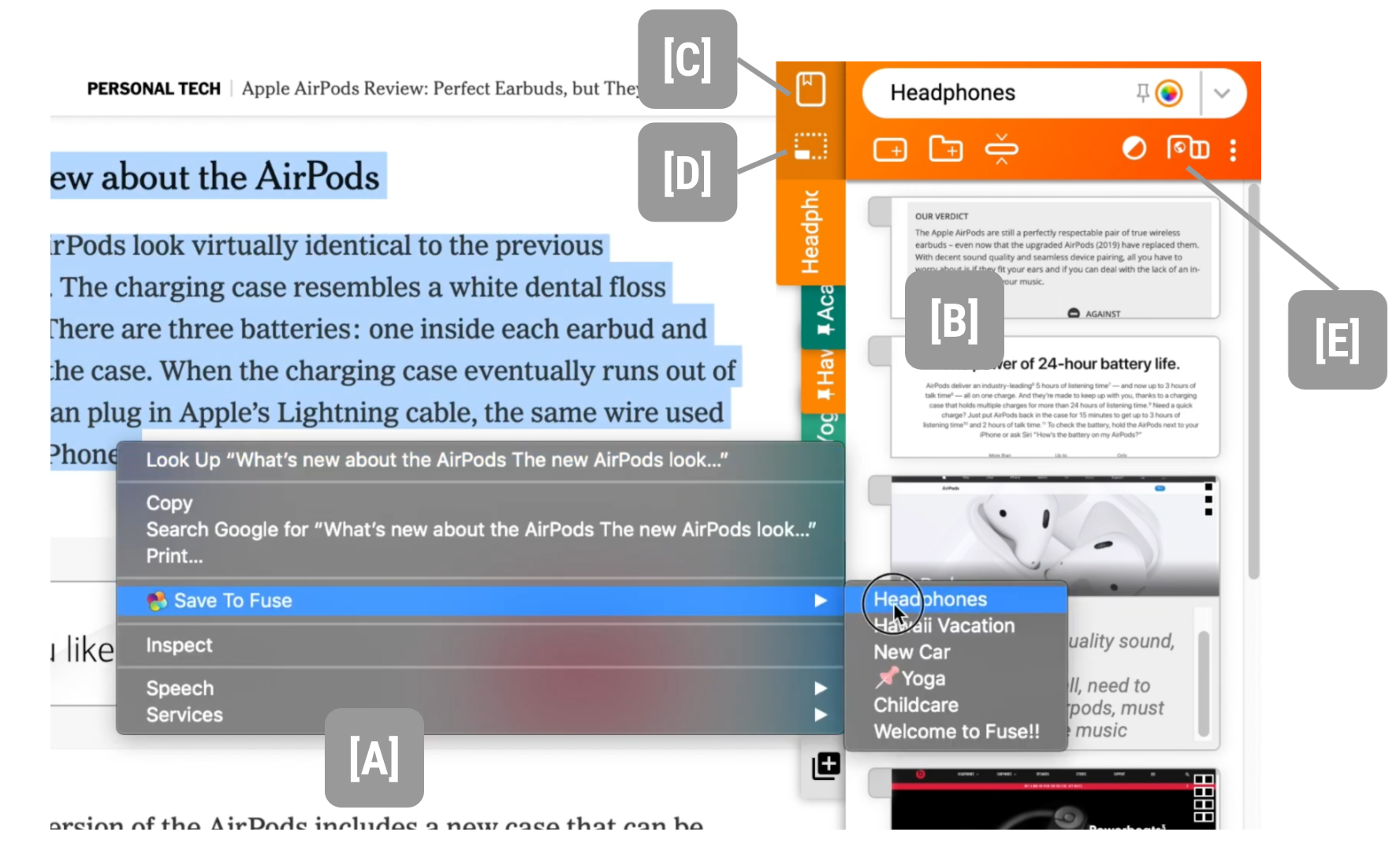}
\caption{The many different ways Fuse enables users to capture web content: (A) Text Highlighting (B) Image Drag and Drop (C) Bookmarking (D) Bounding Box Clipping (E) Tab Importing. 
}
\end{figure*}

\begin{itemize}
    \item \textit{Text Highlighting} \\
    Fuse implements a traditional text highlight to enable users to make high-precision selections. After selection, users can use the context menu to navigate to 'Collect with Fuse' and then select which project to collect into. 
    \item \textit{Image Drag and Drop} \\
    Preliminary user tests revealed that although selecting images was common, doing so with the mouse was much more tedious than dragging. Images that are dragged into a Fuse project will automatically be imported as a card to facilitate easy import of images.
    \item \textit{Bookmarking} \\
    Fuse implements the familiar bookmarking feature, accessible via the sidebar. Doing so saves the entirety of the page, the title and url, and captures the current viewport to be used as the header image of the respective card. 
    \item \textit{Bounding Box Clipping} \\ 
    To facilitate capturing portions of webpages that are neither text nor images, Fuse implements a bounding box clipping tool. Clicking on the respective item in the sidebar triggers a bounding box selection, upon the completion of which will become the header image of the new card. Url and page title data are extracted automatically and associated with the new card. 
    \item \textit{Tab Importing}  \\
    To assist users that would like to capture multiple pages consecutively (as is often the case when closing a browser window), Fuse implements the ability to import all currently open tables into the current Fuse project. Doing so mimics activating the bookmarking capture interaction on every open tab.
\end{itemize}

\subsubsection{[D2] In-situ organizing}

A key design goal we aimed to explore was to enable users to view and organize information as they forage for information, reducing switching costs which produces considerable overhead for users conducting online research \cite{chang2020mesh}.  With this in mind we set out to design an always available sidebar panel that would enable users to create, manage and switch between multiple projects. To do this within the Fuse sidebar, we initially created a menu bar with individual tabs to switch between projects, similar to file folder tabs. Early user-testing of prototypes demonstrated the need to swiftly create and pin new projects, inspiring the creation of two dedicated buttons on the panel for doing so. To better keep users in flow, a user can minimize the Fuse sidebar entirely and reopen it with a keyboard shortcut when they’d like to refer back to it. 

\begin{figure*}[h]
\centering
\includegraphics[width=\textwidth]{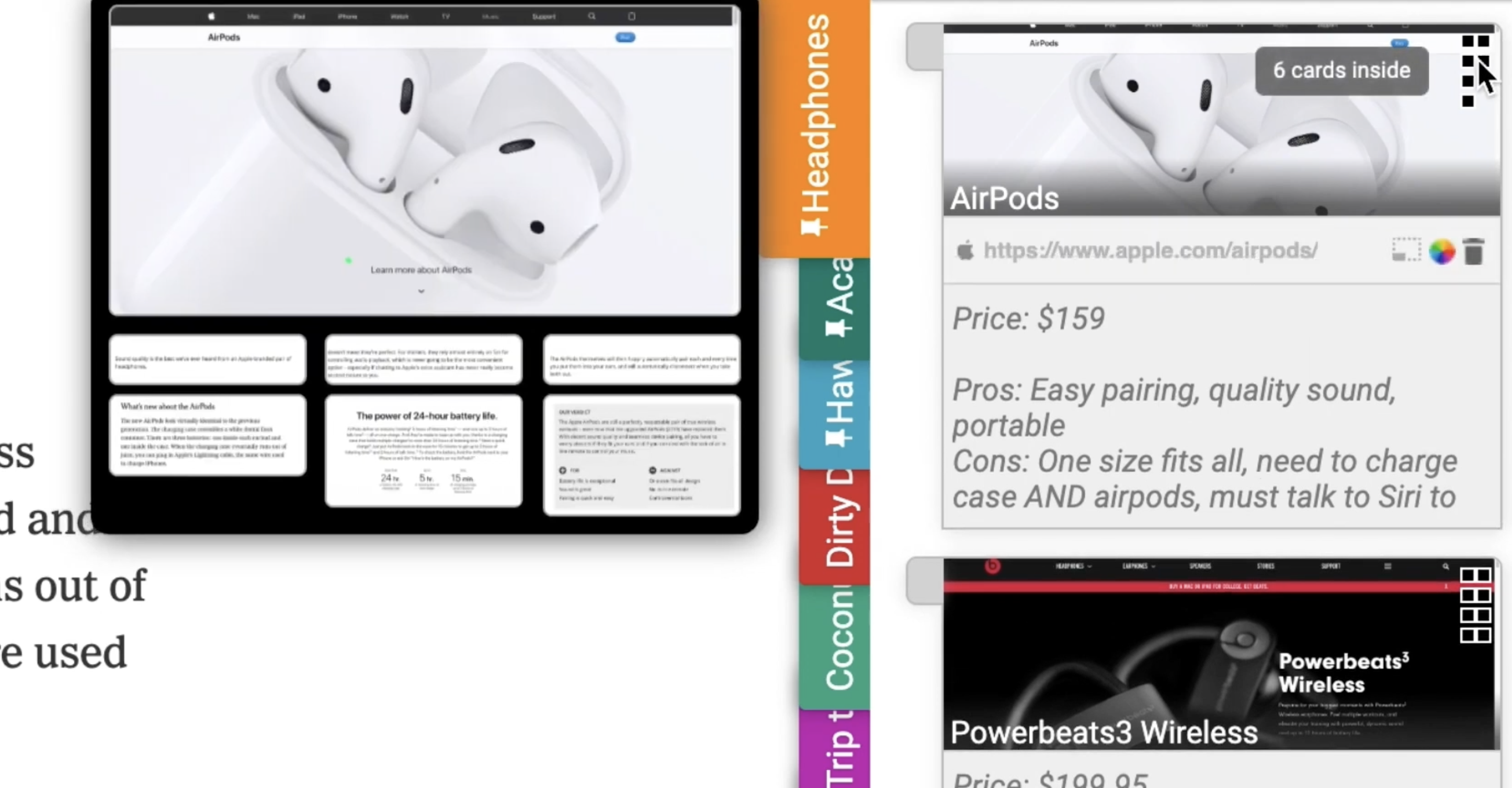}
\caption{The 'peek' interaction within Fuse, which enables users to see cards within a container card without using the 'expanded reader' view. Preview versions of the content appear together in a pop-up left of the sidebar when hovering the mouse over the card preview icons (positioned in the top-right of container cards). 
}
\end{figure*}

Additionally, we aimed for Fuse to support a variety of organizational paradigms in-situ so that users could iteratively organize their collection to best suit the task at hand, as well as account for personal organizational preferences and strategies \cite{fisher2012distributed}. Fuse achieves this in several ways. First, Fuse enables users to use existing cards as 'container cards', which we elaborate on in the next section. Users can also create more traditional folder cards, which are intended to act primarily as labels for explicit hierarchical structures, much like folders in a file system. Internally, these are implemented as standard content cards without the ability to bundle content within them. Lastly, users can utilize the Fuse interface to manually create cards from scratch and assign them text titles and annotations. Users often used these 'manually created cards' as a third type of organization as they could both be used as a simple label and contain bundled content. There are several ways to import content into an organization, regardless of whether it is going into a folder structure or bundled into a container card. Users can drag content such as images from the browsing pane into an existing folder or container card in the Fuse sidebar, automatically creating a new card that will be placed within. Alternatively, users can also drag existing cards from folder to folder or container card to container card in the current collection.


\subsubsection{[D3] Visually compressing items}

Working within the constraints of an always-present sidebar presents significant challenges to representing content and interactions in a compact way that functions across a variety of types of web content, meaning an in-situ system will need to offer a standardized but compact representations of each collected item within the sidebar regardless of the granularity or media type of the content collected. In order to design for this, we decided to represent each piece of collected content using the Card UI paradigm, as it would give users a highly visual representation that could be dragged around to be reordered, and nested. By default, each card would show a header preview image, with a title, source url, and blank annotation. In choosing to do so, we also decided on several transformations so that each type of media imported into Fuse would have a uniform presentation. Bookmarked websites would receive their header images by way of an automatic viewport screenshot during collection, and collected snippets appear directly in the header images. To better support skimming, this header image can be collapsed by clicking. Lastly, Fuse implements a generic 'expanded reader view' (shown in Figure 2), when fully seeing the content and context of an item is desired. In their compressed forms, Fuse cards enable the enable the sidebar to fit many cards and afford the opportunity for us to implement in-situ organization, which we discuss next.

Early user-testing with Fuse revealed that users desired the ability to create structures with multiple nested folders as they used the tool in-situ, necessitating the creation of a compact representation for nested cards within the sidebar. Later iterations of Fuse implement the ability to drag cards into cards, particularly useful when users were collecting many snippets about a singular item such as a product. From this iteration forward, the 'container card' was created as a concept independent of folder cards. 

As opposed to creating hierarchical structures with folders, 'container cards' are primarily meant to simplify structures by compressing many independent pieces of evidence into one card. For example, early user-testing participants often found themselves collecting individual images (such as food images for a restaurant) and nesting them under individual folders named after each option (in this case a restaurant). Instead, Fuse users can nest the many different images within the original, rich card representing the restaurant, creating a bundle of images. Fuse implements several additional affordances to make viewing and previewing the bundled items easier. To view the items, we created a reader view that would pop open on click, showing all of the snippets nested in the card. When not in this view, we implemented a small preview panel that uses a grid of squares to represent the number of snippets (and 'container cards') nested within a specific item to provide a quick overview for the user. Lastly, hovering over this grid of squares activates a small pop-up visual 'peek', with miniaturized versions of each snippet appearing together in a black bubble left of the card being peeked. This interaction can be seen in Figure 4.

\begin{figure*}[h]
\centering
\includegraphics[width=\textwidth]{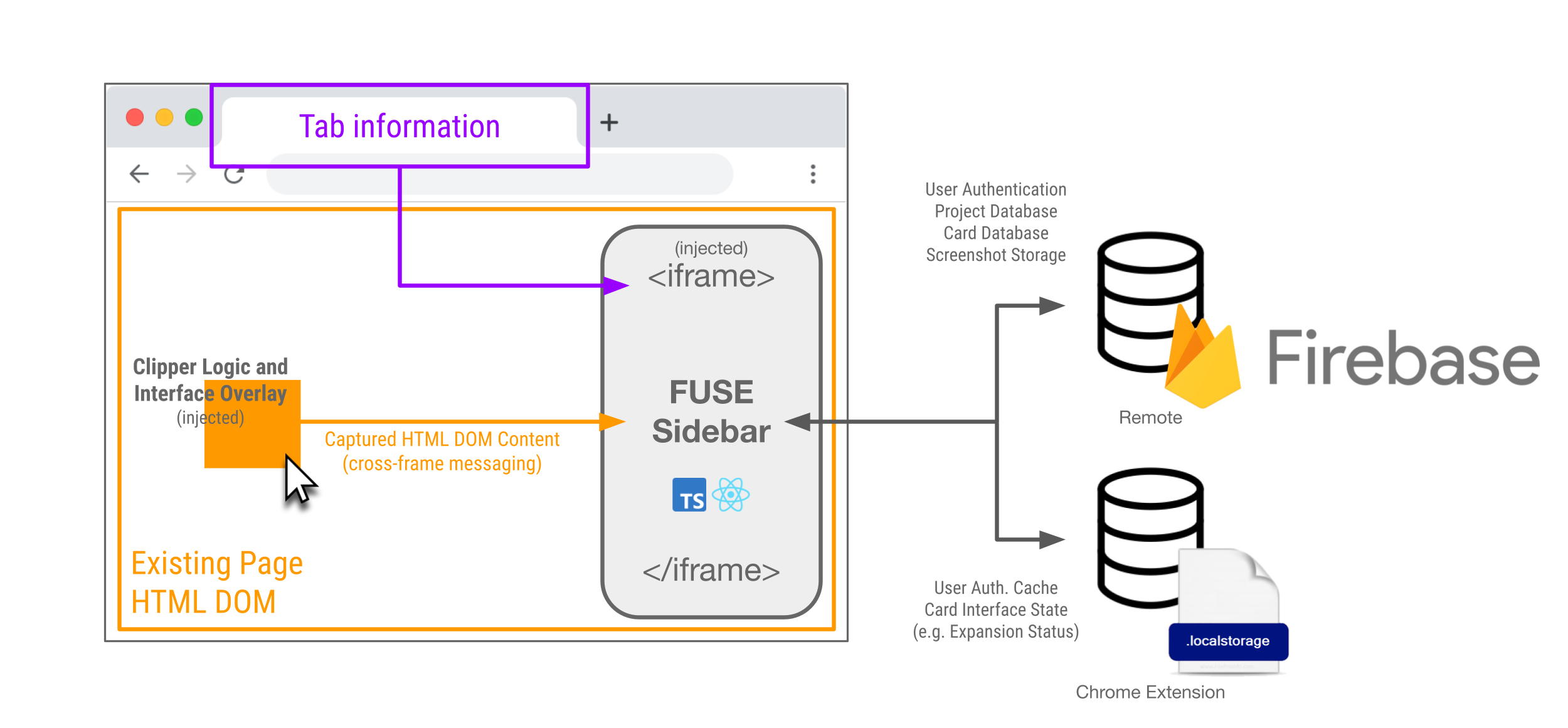}
\caption{Fuse Architecture Diagram. The Fuse sidebar is implemented by injecting javascript code into an iframe attached to the current tab's DOM - enabling the sidebar to capture page content, render the clipper interface overlay, and communicate with both the Chrome user's local storage and the remote Firebase server.
}
\end{figure*}

\subsection{Implementation Details}

Fuse was implemented in approximately 37,000 lines of TypeScript. The UI components were built with the React library and database and authentication implemented with Google Firebase. This enabled users to access their projects across consecutive sessions. The system was implemented as a Chrome extension, which was crucial to keeping several interface interactions as native to the browsing experience. In doing so, we were able to allow users to seamlessly toggle and maintain the Fuse application alongside their normal web foraging behavior with a click or keypress. Second, having read access to website content was critical for implementing the ability for users to clip content on their screen after activating the snipper in the sidebar. Lastly, implementing Fuse as a chrome extension gave the system access to tab information such as url titles and favicons. The sidebar is implemented by attaching an iframe to the current tab’s DOM and embedded onto the page when the extension icon is clicked by the user. Javascript code was injected into the iframe to render Fuse’s interface and logic. This allows us to implement the sidebar in-situ while encapsulating our CSS stylings and program logic. Each iframe in modern browsers also has its own Javascript thread, this design also allows Fuse to run independently of the current tab so the impact on its rendering and performance is minimized. A system architecture diagram can be found in the below figure.

\section{Field Deployment Study}

To better understand how our system can support sensemaking behaviors in online research and  evaluate our prototype, we conducted a field deployment with participants in the wild performing their everyday tasks. 134 distinct users organically participated over a twenty two month period, recruited through a combination of recruiting from authors’ social media feeds, a public email list available on an informational landing page (https://getfuse.io/), and a publicly searchable listing on the Google Chrome app store. The social media posts were short and asked for participants that would “like a flexible tool to help organize online researching” and contained a link to the application’s listing on the Chrome App Store.  Upon installing the application, participants were informed that the application was part of a research study and consent was obtained for the collection of project and card information. Users were not given any specific directions during installation or tasks to perform - the recruitment page consisted of informative content regarding Fuse's collection and organization features, and invited readers to freely download and use the application for their personal online research needs. 

Of these users, 89 (66\%) created projects with non-zero amounts of content. Follow-up interviews were solicited by emailing the top ten most active users in our deployment study (by event count), with eight users responding and participating in a video call follow-up interview, compensated at \$15/hour. As we recruited interview participants, we found significant overlap between themes as we interviewed users such that we were not finding new themes and were reaching saturation with eight participants. In looking at the data across all participants these themes were well reflected in the projects that we inspected or analyzed. Questions spanned several topics detailing participant usage of the application, user utility, the comparative strengths and weaknesses of Fuse (as compared to other applications the user has tried), and areas of potential improvement. The full text of these questions may be found in the supplemental material at the following \href{https://drive.google.com/drive/u/4/folders/1ID87GHiceQrGUnV0Xf8g_jaCFED0qIXU}{link}. 

\section{Results} 

Overall, participants found Fuse to be helpful in visually organizing their online research and used it for both personal tasks such as shopping as well as professional tasks such as academic literature review. One user reported switching over from their browser’s bookmarking interface entirely, using Fuse for organizing their online research throughout the entire 22 month deployment period. As participants were not compensated or recruited for their usage of the application beyond their own utility, we see this as an encouraging sign that Fuse provided sustained value in users’ daily lives and replacing existing tools for these long-term users. Below we discuss themes from two passes of interview coding (one open and one closed) in order to provide an in-depth understanding of how Fuse users derived value from Fuse's designed affordances as they conducted online research during the deployment study. We finish this section with an analysis of the different types of Fuse projects we observed, broken down by user intentions with another pass of coding looking at project and card content.


\subsection{[D1] Collecting content while encoding provenance and context} 

Fuse's first design goal was to lower the friction of collecting information and externalizing their thoughts about that information while conducting online research. In our user interviews, we found evidence that the designed features helped users to import content, augment it with both automatically and manually collected provenence-related information, and organize it into a coherent structure.  

Fuse users described how the extension enabled them to quickly collect key content from the web with lower friction, enabling them to clip a variety of information while avoiding context switching to other tabs or applications. In follow-up interviews, participants were excited by the speed with which Fuse allowed them to import content from their screen into the sidebar: 

\begin{displayquote}
    "[I] can effortlessly throw something in, can easily come back to it later .... Esp if I’m doing shopping online, easy to [throw] stuff in" - P5
\end{displayquote}

\begin{displayquote}
    "[I] did like import tabs - would like to drag into a folder within a project" - P6
\end{displayquote}

In addition to lightweight collection, Fuse users seemed to value the ability to immediately annotate content using the sidebar, best illustrated by this user's remarks:

\begin{displayquote}
    "[I enjoyed] capturing a web page and immediately going in and taking notes about what’s of interest" -P6
\end{displayquote}

Beyond collecting content, Fuse users also seemed to greatly value the ability to embed context within Fuse cards. Previously, provenance has been thought of mostly as the context of where something comes from, but in this latter case, we saw that Fuse enabled users to expand the concept of provenance to content that they needed in order to recollect their own mental context, e.g., why it was chosen, what they were doing, what they still need to do, etc. Within Fuse, this information was collected in a number of ways, both automatically by the Fuse system and explicitly by the users.  Users seemed to value Fuse's automatic collection of basic provenance information, as it allowed them to continue collecting without slowing down to record descriptions, source details, and general impressions:

\begin{displayquote}
"[I] used to track everything in a google doc, [and] had to write a description - taking time .... [I] started using Fuse to have an immediate capture [with the same information]” - P3
\end{displayquote}


 \begin{displayquote}
“[Fuse is nice because] sometimes I screenshot recipes and save it but I forget about it or forget what I named it, also with a screenshot I can never find the source” -P2
\end{displayquote}

\begin{displayquote}
"I like that I can see a little image of the page [when I looked at it later] - helps remind me of what it is" -P5
\end{displayquote}




%

Many Fuse users also manually created annotations for their items; annotations were ubiquitously present across users that created projects with a non-zero amount of content (49\%) and their projects (41\%). Looking more deeply at these annotations, we found that annotations were a key affordance for storing provenance-related information for users looking to store context while conducting a task. To investigate how users did so, we conducted a qualitative code (two passes, one open and one closed) of our annotations of user projects in an attempt to understand what types of information users were collecting. Reviewing all annotations, several things stood out. First, although many note-taking applications build annotation features for attaching free-form text comments to items, we noticed that user annotations in our system primarily contained text extracted from the document, (e.g. the abstract of a research article, highlighting key phrase in a recipe) and user-written summaries of content, with one participant in our interviews (P3) explicitly requesting features around automatic summarization and crowdsourced 'content of importance'. It's likely that key phrases, abstracts, and summaries stood out to Fuse users as valuable annotations because they represent a summary of the collected content that assisted users in overviewing their online research collection. In comparison, few annotations recorded user opinions and comments (e.g. "I do not like this jacket") or the state of the card within some sort of task context (e.g. "looked into this"). 


Second, we looked at the length and size of user annotations, wondering if there was any connection between the media type of content users collected and the amount of user-embedded context they attached to it. We found that user annotations are consistently logarithmic in size and shorter than two sentences across the different types of card content (consistent with prior work regarding user comments (e.g. \cite{horvath2021understanding}, \cite{marshall2004exploring})) with the exception of manually created cards and text snippets. We hypothesize that cards containing text snippets may result from precise extractions that create smaller, 'self-explanatory' content that the user does not feel the need to attach additional provenance content to. Manually created cards seemed to have the largest diversity of (and largest average) lengths, likely because they were less focused on summarizing extracted content and more focused on externalizing user thinking, as indicated by P3: "When I’m [manually] adding a new card, I write my own title and text so I can keep notes in my head". This would be consistent with prior work highlighting the diversity of different strategies users prefer for encoding provenance information and indicates that similar tools may benefit from additional input affordances \cite{fisher2012distributed}.

\begin{figure*}[h]
\centering
\includegraphics[width=\textwidth]{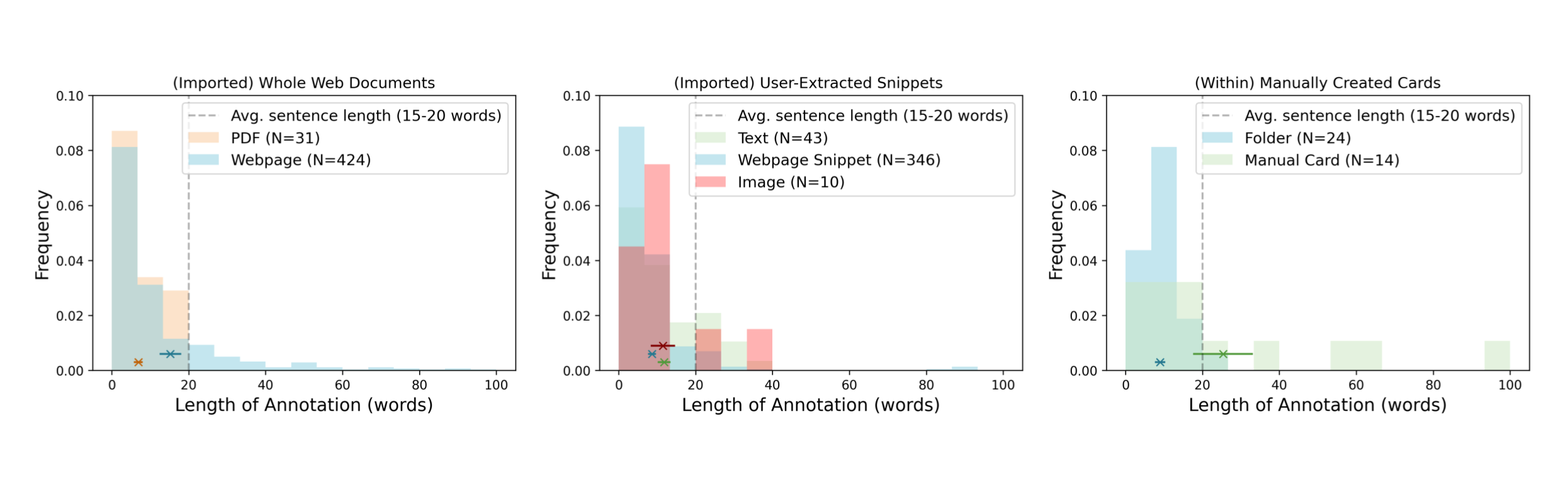}
\vspace{-5mm}
\caption{Bars graph representing the average lengths of annotations by word in different types of Fuse content. The 'x' marks refer to category means and the lines refer to standard error bars (both colored respective to category).  
}
\end{figure*}

Lastly, we investigated if the presence of annotations may be connected with the organizational role of the card in the high-level structure of the collection. Based on our findings regarding users's preference for using annotations for summaries, we hypothesized that the presence of annotation may be linked to a higher position of the card in the structure, for example by summarizing the content of a high-level folder. On the other hand, it's possible that the opposite were true - we knew from earlier that users valued the ability to extract and highlight specific bits of content with annotations, which implied that annotations would be most frequent in leaf node cards that represent low-level evidence, such as an image of a restaurant menu. Instead we found that cards in all positions of hierarchies seemed to be equally likely to have annotations (approximately 16\% of the time), regardless of the whether the card was positioned as a root node within a hierarchy, had content nested inside of it, or was a piece of content nested within another piece of content (a leaf node, functionally).




\subsection{[D2] In-situ organizing
}

Fuse's second design goal focused on the need to enable users to organize information without switching between browsing and a reference document, inspiring the implementation of compact organizational features within Fuse and ultimately the creation of the 'container card' paradigm. Fuse users seemed to value a variety of Fuse's organizational features to help them swiftly organize their content into a coherent structure where things fit together. Nearly all users reordered their collection using the Fuse sidebar (83\%) to ensure a specific order to the cards. The ability to color-code cards was in contrast much less used in practice -  although color was highlighted by a interview participant (P1) as a beloved feature, less than 1\% of Fuse projects used the feature. However, Fuse users seemed able and interested in creating hierarchical structures to compress their collections, with 42\% of users creating hierarchy in their collections. These hierarchies were consistently shallow, with only 4\% of projects containing a structure deeper than 2 levels. Users creating hierarchies did so with a variety of methods, relying about equally on folder cards (46\%) and Fuse's unique "container-cards" (50\%). Users seemed to understand and enjoy the card-in-card paradigm and it's related 'drop' interaction for it's intuitive usage: 

 \begin{displayquote}
"I like the fact that you can organize [the cards and] ...  drop all cards on top of each other (container cards)"-P1
\end{displayquote}

Overall, this new feature was accepted by users, with 50\% using it at least once. In fact, users seemed to prefer this nesting affordance for different types of content, card-in-cards largely held snippet content whereas folders most often were used for holding more traditional bookmarks such as entire webpages or PDFs. We see this as evidence that the card-in-card paradigm was preferable to Fuse users compared to more traditional folders for aggregating web snippets. 

A detailed analysis of the types of organization and projects created by users can be found in the 'Understanding Usage' section below. 

\subsection{[D3] Visually compressing items
}

Fuse's third design goal focused on the need to enable users to visually compress items, resulting in the creation of a compact card UI design and visually compressed representations of nested cards. We found that users valued these features greatly, particularly in their combined ability to provide a visual representation of user's information foraging collections. Users mentioned the impact of this affordance in many ways, whether it was helping their collections feel readily-available for review, be quickly skimmable, and provide the necessary depth to resume tasks. We discuss these themes below.

The potency of having immediate visual access to their entire collection regardless of media type or organizational structure was a common theme in interviews, where users often compared it to current desktop and web paradigms they use:  

\begin{displayquote}
"Fuse seems to me to fit in between [Keep and Evernote] - much lighter weight and right there where you need it" -P7
\end{displayquote}

\begin{displayquote}
“Before with illustrator I have to save the file somewhere but with Fuse ... and I can access it there without having to dig through the files” -P2
\end{displayquote}

\begin{displayquote}
“[I like Fuse] because of the pop out panel and being able to organize things right there and see what you already have … evernote does have a clipper but if you want to look or organize you have to go somewhere else.” -P7
\end{displayquote}

\begin{displayquote}
"[It is] useful to have a quick way of seeing the things I’ve saved, instead of having to go onto Ikea site to check on my shopping cart" -P5
\end{displayquote}

One user also appreciated the ability to toggle the entire sidebar for a more dedicated reading experience, as demonstrated by their user quote:

\begin{displayquote}
"I’m not completely comfortable with something cluttering up the right side of my browser - I would open it if I wanted to look at what’s there, but if I’m browsing I don’t really need it" - P5
\end{displayquote}

Beyond the ability to quickly access their collection, the responses of several users indicated that the Fuse sidebar was valuable for overviewing collections. This activity seemed to have two components. First, Fuse users seemed to enjoy that the collapsed Fuse cards enabled them to see their entire collection at once:   

\begin{displayquote}
With Fuse I can see the thumbnail and see what furniture it is, and can see every folder at once ... Fuse is more accessible [than other tools] because [I] can open it on the browser, can see all the information at once - P2
\end{displayquote}

\begin{displayquote}
I like to read my notes and also scroll all of the materials. - P1
\end{displayquote}

Second, Fuse users found this visual overview to be useful to think about their foraging collections in spatial terms, as illustrated in this user quote: 

\begin{displayquote}
“I like how it works visually, [especially] once I have the tabs established, I can really see where things are ... [I will] soon have 115 articles and poems coming at me - Fuse could be a good way to organize all that - could visualize how to neaten things up” -P4
\end{displayquote}

Lastly, Fuse users indicated that the tool had assisted them in task resumption. For some, the simple presence of the content within the sidebar was helpful as a "constant reminder" (P6). For others, the ability to recover provenance allowed them to quickly jump back into their task and resume it given limited time: 

\begin{displayquote}
"[I] really like having all the information organized for me, [it] helps me when i’m scrolling through what I’ve saved, having the title, having the image there too helps me a lot. Sometimes if I just have 5 minutes I’ll click on things I didn’t have time to read yet" -P3
\end{displayquote}

\begin{displayquote}
"[Fuse is] visually uncluttered and easy to organize information, easier to access and navigate and jog memory compared to bookmarks" - P8
\end{displayquote}

Furthermore, by allowing users to quickly recollect their mental models, Fuse enabled users to think about their tasks in terms of cycles of saving and reviewing, a point made clear in several user responses regarding why they enjoyed Fuse:  

\begin{displayquote}
[I enjoyed Fuse because I] can effortlessly throw something in, can easily come back to it later - P5
\end{displayquote}

\begin{displayquote}
[I enjoyed Fuse because I was] able to very quickly make a bunch of bookmarks - haven’t gone back and organized them too much, because it’s only been a couple of days -P7
\end{displayquote}

\begin{displayquote}
[I enjoyed Fuse to] "organize pages I need to save and revisit/reread" - P8
\end{displayquote}

In summary, Fuse users had found the visual nature of the sidebar system to be useful for a number of activities related to managing their foraging collections, from quickly reviewing their collections to spatially processing their collections when overviewing them all together. These benefits seemed to culminate in the ability to help start and stop foraging tasks, giving users confidence to collect large amounts of content to review later. We discuss the nature of these in-the-wild foraging tasks more in depth in the next section.

\subsection{Understanding Usage}

To further understand the foraging tasks that users were using Fuse for, the primary author went through all 455 projects with their respective cards and annotations in two passes (one open, one closed) to generate project and domain categories from the collections until clear themes emerged \cite{charmaz2019thinking}. In total, 134 users had created Fuse projects over the course of the deployment. Projects with less than three non-folder cards were excluded from the analysis as we found it difficult to infer user intention for projects with very small collections of items, leaving 248 projects. 

At a high level, follow-up interviews had made it clear that users were utilizing Fuse for a diversity of tasks, from organizing instructional materials to comparison shopping and trip planning:

\begin{displayquote}
“[I’ll] soon have 115 articles and poems coming at me - Fuse could be a good way to organize all that - could visualize how to neaten things up” -P4
\end{displayquote}

\begin{displayquote}
“For quick comparison shopping … having [all of my cards] all saved somewhere together is useful …. [also I’m] planning road trip with family next summer - [I am] quickly bookmarking and tracking things we’re thinking about and getting that organized” - P7
\end{displayquote}

\begin{table*}[]
\begin{tabular}{|cccc|ccc|ccccc|c|c|}
\hline
\multicolumn{4}{|c|}{\textbf{T.M.* (14\%)}} & \multicolumn{3}{c|}{\textbf{U.S. (59\%)}} & \multicolumn{5}{c|}{\textbf{C. (23\%)}} & \textbf{S.F. (40\%)} & \textbf{Q.A. (35\%)} \\ \hline
TD & ST & Q & OQ & CO & CL & SY & CO & FE & C/J & EV & DE & - & - \\
54\% & 23\% & 54\% & 14\% & 100\% & 55\% & 4\% & 100\% & 34\% & 14\% & 17\% & 14\% & 100\% & 100\% \\ \hline
\end{tabular}
\vspace{3mm}
\caption{Project Categories based on inferred user intention found in FUSE projects during the field study. Project categories are nonexclusive and as follows: Task Management (TM), Understanding a Subject (US), Comparison (C), Self-Reference (SF), Quick Access (QA). Sublabels are computed within each Project Category and are as follows (from left to right): Todo (TD), State (ST), Queue (Q), Open Question (OQ); Collection (CO), Clipping (CL), Synthesis (SY); Collection (CO), Feature Extraction (FE), Criteria/Judgement (C/J), Evidence (EV), Dependencies (DE). *Not a Project Category, records presence of task management within the collection.}
\label{tab:caption}
\end{table*}

Furthermore, it was clear that users were getting different utility from Fuse for different tasks, for example, some users were only interested in a specific media type: 

\begin{displayquote}
“[Fuse is] good for things that you think look cool - not about bookmarking sites or pages, but more of a repository of nice images” - P5
\end{displayquote}

Our qualitative analysis surfaced five main user intentions within projects. This analysis also generated fifteen distinct domain categories, which we discuss in relation to the user intentions in the section below:

\subsubsection{Understanding a subject}
A common user intention among analyzed projects was to collect resources to research a subject, such as understanding the space of payment processing or conducting a literature review of academic research. These projects were primarily characterized by users who collected multiple documents of information about the one subject or overlapping subjects, extracted clips from the articles, and then (rarely but occasionally; 4\% of the time) synthesized the clips in some sort of external resource such as Google Docs. It is highly likely that the reason we did not encounter many examples of external synthesis stems from the fact that Fuse filled that role for users, diminishing the need to store such links in Fuse projects. The percentage of projects that showed evidence of external synthesis was low, with only about 2.4\% of all projects doing so. However, annotations were frequent additions. We found this user intention in a total of 58.9\% percentage of all projects. 

\subsubsection{Comparing items} 
Several prior systems such as Mesh \cite{chang2020mesh} and Unakite \cite{liu2019unakite} focus on supporting users sensemaking as part of a making a decision. It was therefore of little surprise that we frequently (23.4\% of all Fuse projects) found users creating projects to compare sets of items or services for purchase, such as selecting a piece of furniture or choosing lodging for a trip. Unlike the previous category, projects in this category were characterized by user collection consisting of many individual options. Additionally, users often used the snippet, image extraction, and annotation features of Fuse to surface features extracted features from content and document judgements (e.g. 'this is too expensive'), with several clever users representing dependencies between objects with the use of nesting. In one such example, a user collected 12 different bed frames, nesting alternative frame colors and sidetables under the item that they wanted to associate the two under. Occasionally, Fuse projects contained elements of both comparing items and understanding a subject. This often involved a collection of educational resources for consumers such as articles on 'how to select a bicycle' or 'types of coffee machines'.

\subsubsection{Building a Self-Reference} 
In our review, we identified a more distinct and more advanced version of understanding a subject. In this category, users created lengthy summaries of documents and frequently extracted of key content into the card annotations as part of curating a self-reference within Fuse. This is perhaps unsurprising, as extracting or summarizing content into the annotations was mentioned several times in our interviews with power-users, with one person suggesting Fuse implement the ability to import an automatic summary of the webpage, and another suggesting the use of crowdsourced summaries for popular webpages. This more robust version of collection was popular in the domains of literature review and instructional content, perhaps owing to the fact that such material is often dense and requires summarization or extraction when embedded in a compact collection such as Fuse. Such projects made up 40.3\% of all Fuse projects analysed. 

\subsubsection{Accessing resources quickly (bookmarking)} 
Prior literature has repeatedly shown that users collect content in bookmarks in order to have rapid access to specific resources \cite{abrams1997people}. Such projects were easily identify-able and frequently contained links to coordination materials/hubs or instructional content, such as informational portals. This category most frequently fit into two usage paradigms: users maintained links to internet resources that they needed access as part of conducting another task, or more rarely, the collection had no evident purpose beyond serving as a collection of links that are pleasant for the user to return to. Such collections rarely had items that were related and most likely were collected for their aesthetic or emotional appeal to the user. Altogether, 34.7\% of all Fuse projects were created by users to have swift access to internet resources. 

\subsubsection{Task Management} 
Although we did not design the system for task management, we noticed that some users were using the organizational features of Fuse to document their progress within a task, for example logging which items had already been ordered and which under consideration within a Fuse project regarding Christmas shopping. 14.1\% of all projects contained aspects of task management in this way. Within these projects, we noticed several distinct behaviors. 54\% of such projects contained annotations with explicit todos written down, unsurprising as past social computing researchers have found todos to be a frequent occurrence in notetaking. A similar number of the projects (54\%) had an explicit queue developed for future processing of items, often implemented as a folder with nested items or a bookmarked listicle. Explicit notes regarding the state of a task (e.g. reaching out to a company for a quote when conducting comparison shopping) were present in 22.8\% of such projects. Lastly, a handful of projects (5; 14.2\% of projects with task management), contained externalized 'open questions' that the user was likely holding to guide future exploration.

\section{Discussion and Future Work}



In this work, we explored the idea of an in-situ online research support tool that affords both collection and organization. Our design was motivated by the need to bridge the gap between collecting and organizing web content and explores an interaction paradigm in which the browser can act as a seamless extension of a user's mental workspace. To implement this vision, we created the Fuse prototype, a collapsible, card-based sidebar that is always available to the user. Fuse allowed users to quickly collect pages or clips of web content, organize and reference their structures in-situ, and create a variety of complex structures to support the synthesis of information snippets for a wide range of tasks. Throughout a 22-month deployment study and follow-up interviews with power-users, we found that Fuse enabled users to utilize their browser as an extension of their mental workspace throughout online research tasks in a variety of domains.


Our results suggest that creating applications that interweave content collection and organization can benefit users looking to synthesize web content as part of online tasks. However, fully supporting users performing online research within the browser will require revisiting several areas of study regarding information management in the context of the browser. As demonstrated by our work, a future system designer will need to consider the personal provenance needs of users as they perform online research, as well as the affordances of the environment where this task-related content will be reviewed and accessed after it is initially foraged and curated. The nature of the ensuing task may not be well aligned with the provenance desires of the user when the data was collected and represented in the previous stage, as prior work shows that task schemas of information differ highly between tasks, as well between users. One potential approach to this would be the study and creation of applications that mediate provenance needs for information representations as they are imported from foraging applications such as Fuse to a more complex organizational tools such as personal knowledge management (PKM) software or applications tuned for presentation and sharing such as slideshow software. An early thread in this direction can be found in the work of Liu et al. regarding the information needs of programmers reusing framework-related information \cite{liu2021reuse}. In the more immediate context of supporting in-situ sensemaking, this research direction can also foster new ideas regarding the support of multiple representations of collected artifacts in information foraging software. As we've seen in Fuse, some familiar concepts such as bookmarking are an enduring phenomena while others representations may be best suited for specific user intentions such as comparison. 

Conversations with Fuse users in our follow-up interviews results implied that Fuse assisted users in expanding their working memory during online research by helping users visualize their information collections. As noted earlier, one possible explanation for this positive response could be that visual cards and spatial organizations are easier for browser users to process, therefore increasing the amount of potential items that can be stored before becoming burdensome. This is supported by prior work in psychology 
\cite{gyselinck2011role} as well as web browser design \cite{robertson1998data}. Understanding the empirical  effects of either could be important for the design of future card-based in-situ sensemaking support tools. In our deployment, Fuse users occasionally collected very large projects that could not be viewable without large amounts of scrolling, limiting the amount of spatial information available to the user at any one time. One fruitful direction of research could potentially explore methods of visualizing such large collections that preserve the benefits of spatial and visual memory even when the majority of items are collapsed beyond the view of the user, such as degree of interest graphs \cite{van2009search}, automatic tab clustering \cite{dipin2019user}, or lightweight interactions for previewing and overviewing \cite{greene2000previews}. 

Another explanation for this positive effect may originate from the design of our cards, which are compact and can be collapsed in order to enable overviews of large collections. Since a user would need to process fewer pieces of information per item as they are performing their organizational tasks, this would hypothetically enable them to build representations that contain more items. When a user does need more information about a specific item, they can use the embedded cues within the card to recall additional details, much like Fuse users used viewport screenshots and urls to help recall details about why a specific page was captured. The high utility from these simple cues implies that it's likely that even with our compact design, Fuse cards may hold more information than is necessary for a user before they decide to perform an additional 'recall' step. It remains an open question to determine how much information needs to exist for users performing common online research tasks to be able to perform the deeper recall of features such as provenance, intentions, or quality. Removing this redundant or extraneous information could pave the way to even more compact representations, and thus supporting an even larger working memory for users. Lastly, the ability to structure cards within cards could have also played a role in enabling users to reduce the amount of information they need to interact with by removing the need for users to generate organizational labels that semantically group all nested elements, a burdensome task \cite{abrams1997people}. Instead, users frequently nested items within other items without inputting a label to the higher-level card (aside from the metadata that Fuse provides from the original site).

Fuse also enabled users to maintain page and snippet collections over long durations of use, traditionally a burdensome task for users of explicit bookmarking lists \cite{kaasten2001integrating}. Perhaps attributing to Fuse's ability to support users' ability to recall information about collected items (including in some cases the item's state within the process of an external task), Fuse users were able to resume their foraging and organizing over many sessions - in one case entirely replacing the native bookmarking features of their browser for daily use. Understanding which features of Fuse beyond annotation were most useful to users for the resumption of their processes of collection and organization is still an open question, although such work would be valuable for any future extensions into the mobile context, where interruption is common and session times are generally shorter. The topic of suspension and resumption of tasks in the browser has been approached by researchers from several angles, such as re-using searches and re-finding \cite{morris2008searchbar} as well as task-focused approaches to grouping browser tabs \cite{rajamanickam2010task} \cite{wang2010multitasking}. This direction of research also has strong parallels with the study of personal task management (PTM) tools and tools that explore the intersection of task management and browser tab organization \cite{chang2021tabsdo}. Leveraging the insights from such work can be key to optimising systems such as Fuse for long-term use, whether it be understanding how to best hide old, irrelevant items or maintaining quick access to useful resources as collections grow.   

\subsection{Beyond Collecting and Curating}
Our interviews with power-users made it clear that their foraging within Fuse was situated alongside a task context that would involve other activities and applications that are specialized such as for accounting or documentation: 

\begin{displayquote}
    "A spreadsheet is pretty good for keeping track of the budget [of Christmas shopping], which I don’t think fuse would be good at" - P7
\end{displayquote}

\begin{displayquote}
"[If possible, I'd like Fuse to help me] create citations/references - would be a lot of value to me" - P6
\end{displayquote}

As previously mentioned, these user quotes suggest that tools aiming to support sensemaking in the browser will likely need to be designed with tasks-specific information representations in mind. This need would be further amplified if Fuse collections were extended to be shareable between users, as it's likely that re-users of Fuse collections will have different goals and need to re-represent the data. The grander vision of Fuse could perhaps take a similar form to that of Vannevar Bush's vision of the Memex, wherein information collections would not just be stored for individual users, but also shared with others for use in exploration \cite{nyce1991memex}. In the context of Fuse and online research, enabling the sharing of information collections in a way that is productive to other users would require an understanding of how and why sensemaking reuse occurs. Additionally, it would be important to understand how users select which parts of other's sensemaking collection to co-opt, and what factors influences this act of trust. Work in this vein is far and few in between, although researchers have studied this in the context of comparison charts for software documentation \cite{liu2021reuse}.

In this work, we explored the idea of an in-situ online research support tool that aims to support seamless collection, externalization, and organization across a wide range of domains. In doing so, we created the Fuse prototype, a collapsible, card-based sidebar that is always available to the user. Throughout a 22-month deployment study and follow-up interviews with power-users, we found that Fuse supported their working memory in online research tasks in a variety of domains. Although Fuse was a prototype system, we hope it spurs the development and exploration of sensemaking support tools that seamlessly combine the experience of browsing and curating to support online research directly into the browser.

\begin{acks}
This work was supported by NSF grants FW-HTF-RL-1928631, SHF-1814826, and the Office of Naval Research. We'd also like to thank the reviewers for their generous review and feedback. Finally, we'd also like to acknowledge our study participants, without whom this work would not have been possible.

\end{acks}

\bibliographystyle{ACM-Reference-Format}
\bibliography{sample-base}

\end{document}